\documentclass[conference]{IEEEtran}
\IEEEoverridecommandlockouts

\usepackage{array}
\usepackage{url}
\usepackage{float}
\usepackage{cite}
\usepackage{amsmath,amssymb,amsfonts}
\usepackage{algorithmic}
\usepackage{graphicx}
\usepackage{textcomp}
\usepackage{listings}
\usepackage{xcolor}

\lstdefinestyle{mystyle}{ 
    commentstyle=\color{gray},
    keywordstyle=\color{magenta},
    numberstyle=\tiny\color{gray},
    stringstyle=\color{blue},
    basicstyle=\ttfamily\footnotesize,
    breakatwhitespace=false,         
    breaklines=true,                 
    captionpos=b,                    
    keepspaces=true,                 
    numbersep=5pt,                  
    showspaces=false,                
    showstringspaces=false,
    showtabs=false,                  
    tabsize=1
}
\def\BibTeX{{\rm B\kern-.05em{\sc i\kern-.025em b}\kern-.08em
    T\kern-.1667em\lower.7ex\hbox{E}\kern-.125emX}}
\begin{document}

\title{Towards Machine-actionable FAIR Digital Objects with a Typing Model that Enables Operations \\
\thanks{This project is funded by the Helmholtz Metadata Collaboration Platform (HMC) and supported by the consortium NFDI-MatWerk, funded by the Deutsche Forschungsgemeinschaft (DFG, German Research Foundation) under the National Research Data Infrastructure – NFDI 38/1 – project number 460247524.}
}
\author{\IEEEauthorblockN{Maximilian Inckmann}
\IEEEauthorblockA{\textit{Scientific Computing Center}\\
\textit{Karlsruhe Institute of Technology}\\
Karlsruhe, Germany\\
ORCiD: 0009-0005-2800-4833}
\and
\IEEEauthorblockN{Nicolas Blumenröhr}
\IEEEauthorblockA{\textit{Scientific Computing Center}\\
\textit{Karlsruhe Institute of Technology}\\
Karlsruhe, Germany\\
ORCiD: 0009-0007-0235-4995}
\and
\IEEEauthorblockN{Rossella Aversa}
\IEEEauthorblockA{\textit{Scientific Computing Center}\\
\textit{Karlsruhe Institute of Technology}\\
Karlsruhe, Germany\\
ORCiD: 0000-0003-2534-0063}}

\maketitle

\begin{abstract}
FAIR Digital Objects support research data management aligned with the FAIR principles.
To be machine-actionable, they must support operations that interact with their contents. 
This can be achieved by associating operations with FAIR-DO data types.
However, current typing models and Data Type Registries lack support for type-associated operations.
In this work, we introduce a typing model that describes type-associated and technology-agnostic FAIR Digital Object Operations in a machine-actionable way, building and improving on the existing concepts.
In addition, we introduce the Integrated Data Type and Operations Registry with Inheritance System, a prototypical implementation of this model that integrates inheritance mechanisms for data types, a rule-based validation system, and the computation of type-operation associations.
Our approach significantly improves the machine-actionability of FAIR Digital Objects, paving the way towards dynamic, interoperable, and reproducible research workflows.
\end{abstract}


\section{Introduction}
\label{sec:Introduction}
The rapid advancement in digital technologies has significantly transformed scientific research, facilitating the collection, processing, and analysis of extensive datasets.
However, the growing diversity and complexity of research data present substantial challenges, particularly in terms of findability, accessibility, interoperability, and reusability~\cite{directorate-generalforresearchandinnovationeuropeancommissionTurningFAIRReality2018}.
To address these challenges, the FAIR principles~\cite{wilkinsonFAIRGuidingPrinciples2016c} were established, guiding best practices for sustainable and efficient management of research data.

FAIR Digital Objects (FAIR-DOs)~\cite{andersFAIRDigitalObject2023, schultesFAIRPrinciplesDigital2019a} embody these principles, aiming to provide common mechanisms to enable machine-actionable, persistent, and harmonized representation of (meta)data beyond the borders of data spaces~\cite{directorate-generalforresearchandinnovationeuropeancommissionTurningFAIRReality2018, blumenrohrFAIRDigitalObjects2024}.
FAIR-DOs use globally unique, resolvable, and persistent identifiers (PIDs) with their persistent records based on the well-established Handle system~\cite{RFC3650}, which ensures their longevity and reliable referencing.
Every value inside a FAIR-DO record is assigned a data type that is always referenced using a PID and defines the syntax as well as the semantic meaning of this value.
This data type should be reused wherever its syntax and semantics fit, ensuring that identical references denote identical meaning.
This generates harmonized artifacts which are interpretable and processable by a machine to, e.g., determine available operations for FAIR-DOs.
Multiple data types may be aggregated in profiles that define the structure of a FAIR-DO record.
Beyond the borders of research domains, domain-agnostic profiles, such as the Helmholtz Kernel Information Profile~\cite{curdtHelmholtzMetadataCollaboration2022}, are used to harmonize essential information in FAIR-DOs.
The strong and manifested type system of FAIR-DOs is therefore the foundation for machines to automatically interact with them and with their referenced resources across research domains~\cite{blumenrohrFAIRDigitalObjects2024,weilandFDOMachineActionability2022}.

FAIR-DO Operations describe a mechanism for type-based interaction with FAIR-DOs and their contents, thus making them machine-actionable.
To allow for automatic execution, they need to be described in a fully typed, interoperable, technology-agnostic, and reusable manner.
To enable the computation of available FAIR-DO Operations for a given FAIR-DO, we need to bidirectionally associate them with data types in a type system.
This leads to a highly inter-connected typing model that needs to be managed, queried, and validated.

Existing Data Type Registries (DTRs)~\cite{schwardmannAutomatedSchemaExtraction2016a} with their typing models represent a significant development towards machine-interpretable FAIR-DOs.
However, their schema-reliant architecture makes them unable to utilize and provide complex mechanisms beyond the capabilities of JSON schema, and thus cannot facilitate type-associated FAIR-DO Operations.
Such capabilities are needed to, e.g., bidirectionally associate FAIR-DO Operations to data types, to realize inheritance mechanisms and to deal with the highly connected typing model that is required for FAIR-DO Operations.

To address these shortcomings, we developed a typing model for a new graph-based FAIR-DO type system that we prototypically implemented as the Integrated Data Type and Operations Registry with Inheritance System (IDORIS) to showcase its feasibility.
This typing model is conceptually based on the components of current DTR systems and lessons learned through their usage.
We leverage the resulting type system to model type-associated FAIR-DO Operations in a technology-agnostic, highly reusable, and well-described manner.
We come up with a comprehensive solution that integrates FAIR-DO Operations as type-associated operations, inheritance, and semantic validation within a single type system.
Hence, this work contributes to the field of FAIR (research) data management by achieving substantial progress in the long-term vision of machine-actionability.

The paper is organized as follows: 
Section~\ref{sec:SOTA} provides a review of relevant technologies and related work.
In Section~\ref{sec:Model}, we describe our typing model as a basis for type-associated FAIR-DO Operations and in Section~\ref{sec:Impl} we introduce IDORIS as a prototypical implementation of this model.
In Section~\ref{sec:Evaluation}, we discuss and evaluate our model as well as the prototypical implementation based on a specific use-case.
Finally, Section~\ref{sec:Conclusions} summarizes the key contributions and discusses future research directions.

\section{Related Work}
\label{sec:SOTA}
FAIR-DOs essentially constitute a data management approach comparable to Linked Data~\cite{bizerLinkedDataStory2009} or nano-publications~\cite{grothAnatomyNanopublication2010}, but are distinguished primarily by their emphasis on persistence and strong type-safety, which ensure their machine-interpretability.
Within the Research Data Alliance (RDA)\footnote{\url{https://rd-alliance.org}}, multiple working groups and interest groups established outcomes and recommendations on FAIR-DO content~\cite{weigelRDARecommendationPID2019}, information typing models~\cite{weigelPIDInformationTypes2015}, and DTRs~\cite{lannomRDADataType2015b}, which have been acknowledged, among others, by the European Commission~\cite{directorate-generalforresearchandinnovationeuropeancommissionTurningFAIRReality2018, europeancommissionCommissionImplementingDecision2017}.
Despite ongoing discussions and early implementations within international initiatives such as the RDA and the FAIR Digital Objects Forum\footnote{\url{https://fairdo.org/}}, comprehensive practical solutions addressing critical gaps in the FAIR-DO typing infrastructure still need to be developed and widely adopted.
The existing typing infrastructure comprises three Data Type Registry instances – the ePIC test DTR\footnote{\url{https://dtr-test.pidconsortium.net}}, the ePIC production DTR\footnote{\url{https://dtr-pit.pidconsortium.net}} and the EOSC DTR\footnote{\url{https://typeregistry.lab.pidconsortium.net}}.
Those DTRs follow the same typing model, with three schemas that constitute the basis for information typing: \textit{PID-BasicInfoTypes}, \textit{PID-InfoTypes} and \textit{KernelInformationProfiles}~\cite{schwardmannAutomatedSchemaExtraction2016a,weigelRDARecommendationPID2019}; 
However, their implementations slightly differ between the DTR instances and are not standardized.
These systems were already used to model information types for FAIR-DOs in the frame of several use cases from different domains, e.g., in material sciences~\cite{avilacalderonManagementReferenceData2025}, in digital humanities~\cite{krausGoldStandardBenchmark2024}, and in energy research~\cite{mayerThermalBridgesBuilding2023}.

Technologically, all current DTRs are based on Cordra~\cite{tupelo-schneckIntroductionCordra2022}, a JSON schema-based metadata repository that is only able to validate the syntactic compliance to JSON schemas.
Thus, the primary focus of existing DTRs has been limited to syntactic validation, offering at most rudimentary support for semantic validation or linking executable operations to specific data types.
Moreover, despite the recognized benefits of object-oriented programming (OOP) principles such as inheritance and polymorphism in software engineering, current implementations of information types by DTRs either completely lack or inadequately support these mechanisms.
This absence of sophisticated logic significantly restricts the semantic richness and operational flexibility needed to represent complex, interconnected data resources commonly encountered in scientific research.

In addition, the DTRs do not support systematic reuse of type definitions, significantly hindering scalability and maintainability.
To enhance the domain-specific expressivity of FAIR-DOs, it can be desired to create a Kernel Information Profile (KIP)~\cite{weigelRDARecommendationPID2019} of selected domain-specific attributes in addition to those provided by the domain-agnostic ones, e.g. Helmholtz KIP~\cite{curdtHelmholtzMetadataCollaboration2022}.
This case is not adequately supported by the existing DTRs, forcing creators of such domain-specific profiles to remodel the specification of the domain-agnostic KIP.
This leads to redundant work, reduced reusability, and is a missed opportunity to leverage the de-facto subtyping relation between domain-specific and -agnostic KIPs.

Conceptually, FAIR-DO Operations provide a mechanism to interact with FAIR-DOs, i.e., the values contained within the Handle records, and the external resources they reference 
(i.e., the bit sequence)~\cite{weilandFDOMachineActionability2022}.
Currently, multiple approaches exist for service-oriented FAIR-DO Operations.
They typically focus on basic CRUD (Create, Read, Update, Delete) functionalities as specified by the Digital Object Interface Protocol (DOIP)~\cite{kahnDigitalObjectInterface2018a}.
They operate at the level of the FAIR-DO as a whole, and must be individually implemented by each service that supports such operations~\cite{kahnDigitalObjectInterface2018a}.
Currently, there is no method to describe technology-agnostic FAIR-DO Operations independently from the specific executing service and to dynamically associate them to FAIR-DOs according to at least one FAIR-DO association mechanism, i.e., ``Record typing'', ``Profile typing'', and ``Attribute typing'', as described in ~\cite{blumenrohrComparativeAnalysisModeling2025}.
\\

These existing limitations highlight the necessity of a more advanced typing infrastructure that is capable of supporting sophisticated semantic validation, inheritance management, polymorphism, and robust type-associated FAIR-DO Operations within FAIR-DO ecosystems.

\section{Typing Model}
\label{sec:Model}
\begin{figure*}[htbp]
\centerline{\includegraphics[width=7in]{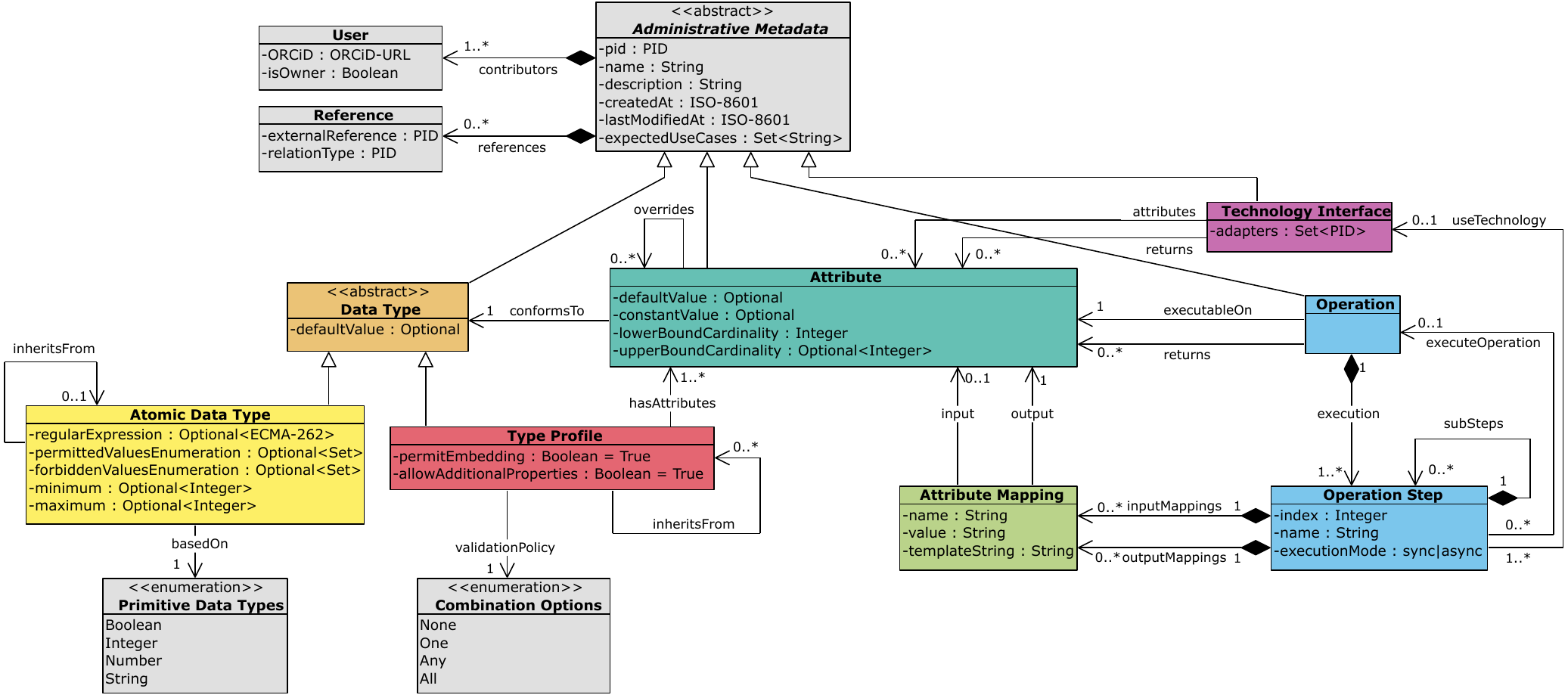}}
\caption{UML model class diagram of the complete typing model}
\label{fig:typing-model}
\end{figure*}

Before going into the specifics of our model, we need to provide a brief overview of the current technical implementation of typing for FAIR-DOs.

Every FAIR-DO is described by an information record of key-value pairs, stored in the Handle Registry\footnote{\url{https://handle.net}} and resolvable by a Handle PID ~\cite{blumenrohrFAIRDigitalObjects2024,RFC3650}.
A key in the information record uses a PID to reference a machine-interpretable information type in a DTR.
This allows the value to be validated against the referenced information type.
We use the term ``typing'' to refer to the availability of information within FAIR-DO records and our typing model.
This is similar to the ``information typing'' used within current DTRs, but opposite to ``FAIR-DO typing'' in the context of association mechanisms for operations as proposed in~\cite{blumenrohrComparativeAnalysisModeling2025}.

On this basis, we introduce our typing model for FAIR-DOs, the description of technology-agnostic FAIR-DO Operations and the association between data types and operations, indicating the analogies to OOP principles.
Figure~\ref{fig:typing-model} depicts the typing model for FAIR-DOs in a colorized UML diagram, consisting of the following classes: \textit{Data Type} as a generalized term (orange), \textit{Atomic Data Type} (yellow), \textit{Type Profile} (red), and \textit{Attribute} (dark green).
Likewise, technology-agnostic FAIR-DO Operations are associated with our typing model through the \textit{Attribute} class and consist of instances of the classes \textit{Operation} and \textit{Operation Step} (blue), \textit{Technology Interface} (purple) and \textit{Attribute Mapping} (light green).
The gray elements are enumerations and administrative metadata that partially depend on the implementation.
For simplicity, we write the names of the classes in lowercase italics to refer to their instances, and in uppercase italics to refer to the classes themselves.

\subsection{Data Types}
\label{subsec:model:DataTypes}
We use Data Type as a generalized term to refer to Atomic Data Types and Type Profiles by specifying the \textit{Data Type} class as an abstract superclass of the \textit{Atomic Data Type} and \textit{Type Profile} classes.
This abstraction allows us to reference \textit{data types} consistently, thereby reducing the redundancy and complexity of our model and its implementation whilst enhancing its semantic clarity and expressivity.
Likewise, as detailed in Subsection~\ref{subsec:model:Attributes}, this also allows the definition and logic of attributes in the \textit{Attribute} class to be agnostic towards the instances of the \textit{Data Type} subclasses they conform to.

\subsubsection{Atomic Data Types}
\label{subsubsec:model:AtomicDataType}
Instances of the \textit{Atomic Data Type} class define the syntax of every value in the information record of any FAIR-DO.
They are built on top of primitive JSON types (Boolean, Integer, Number, or String) to enable JSON serialization.
Therefore, \textit{atomic data types} are comparable to primitive data types in OOP, but offer additional restriction mechanisms that allow for a more strict validation of values:
for any \textit{atomic data type}, predefined constant enumerations of permitted and forbidden values can be specified, which are prioritized over the following mechanisms.
Strings can be limited by specifying a regular expression, as well as a minimum and maximum length.
Integers and decimal numbers can be limited by providing a minimal and maximal value.
These restrictions of the value space guarantee the quality and syntactic correctness of the information contained in FAIR-DOs, which benefits machine-interpretability.

To make \textit{atomic data types} and their potential association with operations reusable and consistent, we introduce a simple hierarchical inheritance mechanism:
they can optionally refer to at most one parent, which is intended to have a broader definition.
Upon validation of a value for an \textit{atomic data type}, this value needs to be correctly validated against all \textit{atomic data types} in the inheritance chain.

\subsubsection{Type Profiles}
\label{subsubsec:model:TypeProfile}
\textit{Type profiles} specify the structure and content of a FAIR-DO by associating a set of typed \textit{attributes} that are instances of the \textit{Attribute} class.
\textit{Attributes} represent \textit{data types} and additional semantics, which will be further explained in Subsection~\ref{subsec:model:Attributes}.
The validation policy determines which combination of \textit{attributes} must be available, and whether to allow or forbid additional \textit{attributes} in a FAIR-DO complying with the \textit{type profile}.
The options are to allow none, exactly one, any but at least one, or all of the attributes.
A \textit{type profile} can describe the entire structure of FAIR-DO records and complex JSON objects that are used as values of a specific \textit{attribute} within a FAIR-DO record.
The latter option is particularly useful when dealing with intricate or tightly-coupled information that is not generic enough to be extracted into a separate FAIR-DO but still needs to be processed together.
For instance, the description of measurement units requires storing a value, a unit, and possibly some information about its accuracy together.

In addition, instances of the \textit{Type Profile} class can make use of a multi-inheritance mechanism.
Despite being known to cause problems such as naming conflicts in programming languages~\cite{martinJavaCriticalComparison1997}, this is not the case in our model since every \textit{data type} and \textit{attribute} is assigned a PID, making them unambiguously addressable.
The remaining potential conflicts of multi-inheritance can be solved through heuristics, whose implementation details are outside the scope of this work.
\textit{Type profiles} are therefore comparable to classes in OOP.

\subsection{Attributes}
\label{subsec:model:Attributes}
An \textit{attribute} points to a \textit{data type} that defines its value space and a default value, if any.
\textit{Attributes} specify their cardinality by providing a lower boundary~\(l\) and optionally an upper boundary~\(u\).
This enables them to represent optional single values (\(l=0;u=1\)), mandatory single values (\(l=1;u=1\)), limited lists (\(l\geq0;u\geq2\)), and unlimited lists (\(l\geq0;u=\perp\)) of values.
\textit{Attributes} behave covariantly when they are used in FAIR-DO information records or as a return value of an \textit{operation} as detailed in~\ref{subsubsec:model:ops:ops}.
Since \textit{attributes} are assigned a PID and contain elements of the \textit{Administrative Metadata} class, they can be referenced directly to specify a value within a FAIR-DO record.
This is necessary in case multiple values that conform to the same \textit{atomic data type} are used in a \textit{type profile}.
For instance, the Helmholtz KIP~\cite{curdtHelmholtzMetadataCollaboration2022} includes ``dateCreated'' and ``dateModified'', both adhering to the ISO 8601 standard, which is represented as an \textit{atomic data type}.
Without directly referencing these \textit{attributes}, both values would refer to the identical PID of the ISO 8601 \textit{atomic data type}, resulting in a loss of valuable semantic differentiation.

This approach to \textit{attributes} resembles object attributes or variables in OOP, both in terms of functionality and semantics.
However, \textit{attributes} according to the \textit{Attribute} class in our model additionally fulfill the crucial role of associating FAIR-DO Operations with \textit{data types}.


\subsection{Modeling FAIR Digital Object Operations}
\label{subsec:model:ops}
In the following, we outline the methodology for modeling technology-agnostic FAIR-DO Operations based on the previously introduced classes of the typing model for FAIR-DOs.
For this, we need to abstract and describe these operations as well as technologies, enrich them with meaningful metadata, and provide a mechanism to adapt this generic definition to actual execution environments in order to enable automatic computation.

\begin{figure*}[htbp]
\centerline{\includegraphics[width=7in]{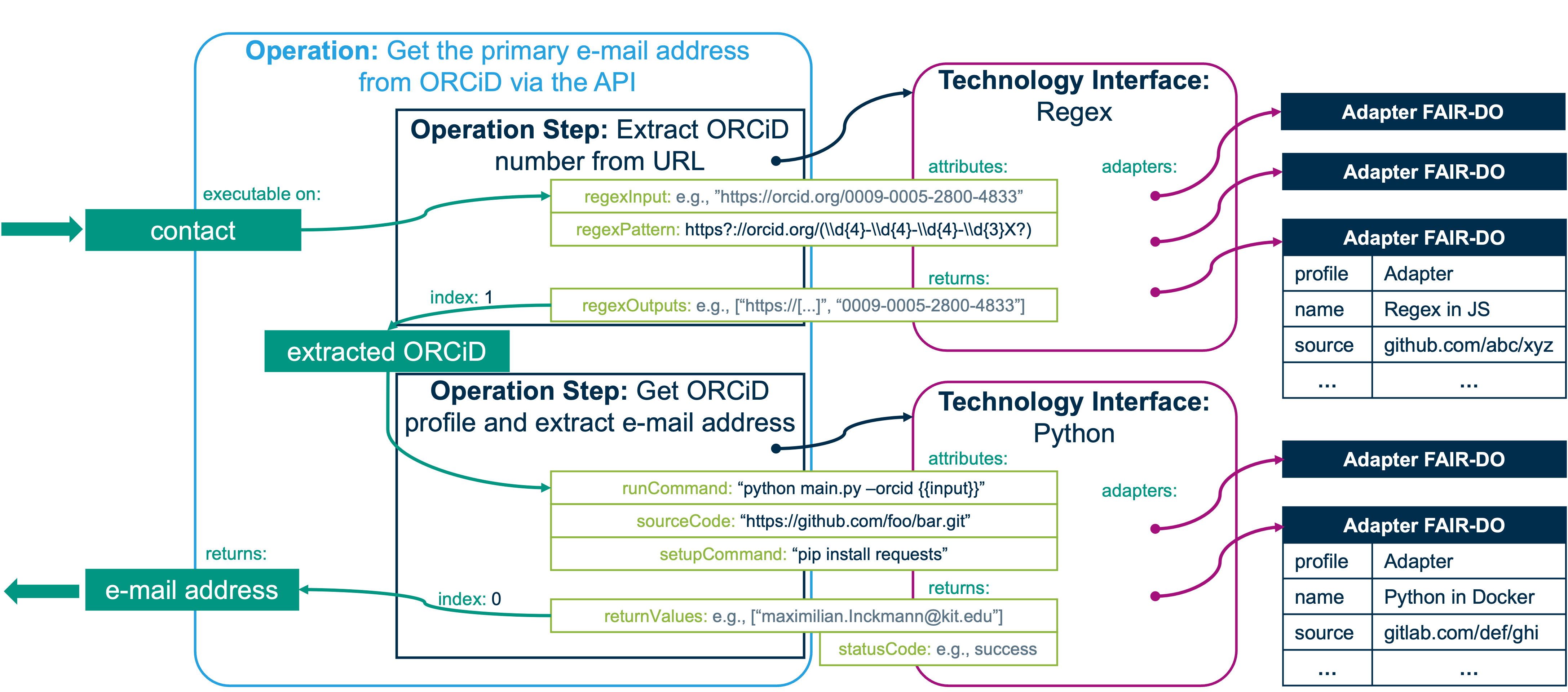}}
\caption{Example of a complete FAIR-DO Operation}
\label{fig:operation-example}
\end{figure*}

Figure~\ref{fig:operation-example} is a visual example of a possible application of a FAIR-DO Operation.
The depicted operation, modeled by instances of the \textit{Operation} and \textit{Operation Step} classes, receives an ORCiD via the ``contact'' \textit{attribute}, contained within the Helmholtz KIP~\cite{curdtHelmholtzMetadataCollaboration2022}, extracts the ORCiD number/letter sequence, and returns the ``primary e-mail address'' of the ORCiD profile as the result.
During the execution, two distinct technologies are used that are modeled by instances of the \textit{Technology Interface} class (described in Subsection~\ref{subsubsec:model:Technology}): a regular expression (Regex) and a Python Script.

\subsubsection{Operations}
\label{subsubsec:model:ops:ops}
The \textit{Operation} class describes an action that can be performed on an instance of the \textit{Attribute} class to which it is applicable.
Therefore, it must reference this \textit{attribute} and all \textit{attributes} it returns.
\textit{Operations} always contain a non-empty ordered list of instances of the \textit{Operation Steps} class (described in Subsection~\ref{subsubsec:model:ops:steps}), that specify all the tasks that are performed during the execution of an \textit{operation}.
When comparing FAIR-DO Operations to OOP, instances of the \textit{Operation} class are similar to both, functions and methods with exactly one input parameter and possibly multiple return values.
They are bound to instances of the \textit{Attribute} class (or the respective \textit{data type}) on which they are executable, thereby resembling methods.
However, as they do not have the capability to directly modify the associated values stored within the FAIR-DOs and are stateless, they align more closely with the typical characteristics of a function.

\subsubsection{Operation Steps}
\label{subsubsec:model:ops:steps}
\textit{Operation steps} can be understood as tasks in an operation workflow.
The order of execution of the \textit{operation steps} within an \textit{operation} is specified in ascending order by the index and the availability of the \textit{attributes}.
An \textit{operation step} specifies whether it uses a technology via an instance of the \textit{Technology Interface} class (Subsection~\ref{subsubsec:model:Technology}), uses another \textit{operation}, or contains multiple \textit{operation steps}.
It also contains a set of input and output mappings that use \textit{attribute mappings} (Subsection~\ref{subsubsec:model:Mappings}) to connect, transform, and specify values between \textit{attributes}, depending on the used \textit{technology interface} or \textit{operation}.
Due to this strong coupling to the scope of the \textit{operations}, \textit{operation steps} and \textit{attribute mappings} are not reusable, not assigned a PID, and managed as composites inside an \textit{operation}.
\textit{Attributes}, \textit{technology interfaces}, and \textit{operations} on the other hand rely heavily on their reusability and are therefore assigned a PID.
\textit{Operation steps} are comparable to function calls on a \textit{technology interface} or another \textit{operation}.
They can, however, also be seen as subroutines.

\subsubsection{Technology Interface}
\label{subsubsec:model:Technology}
Due to the high effort invested in specifying, testing, and executing a technology, it is desirable to make this work highly reusable.
We therefore decided to separate the technologies from the problem-specific operations.
Instances of the \textit{Technology Interface} class realize the reusability layer by providing an environment-independent interface to the execution.
\textit{Technology interfaces} specify a set of input \textit{attributes}, a set of output \textit{attributes}, and reference a set of Adapter FAIR-DOs via their PIDs.

These Adapter FAIR-DOs are specific to the executing environment and actually implement how the \textit{technology interface} is executed on any given system.
In our envisioned approach, the executing systems specify a \textit{type profile} for their \textit{adapters}, which in turn specify machine-interpretable information.
This enables the adapter to be downloaded, verified in its integrity, and executed, being subject to the security policies of the executing system.
We did not model Adapter FAIR-DOs as classes in our model (Figure~\ref{fig:typing-model}) since they just need to be uni-directionally referenced from the \textit{technology interfaces} and their content might vary between the executing systems which is perfectly solved by \textit{type profiles} in FAIR-DO records that also have a PID.

Relating this to the regular expressions used in our example (Figure~\ref{fig:operation-example}), we see that there are different libraries and APIs for different environments.
In this case, we propose to create a \textit{technology interface} for the ``Regex'' technology that accepts an input string and a pattern while returning an array of strings for the regex groups (the first element always represents the fully validated string).
For this ``Regex'' technology, we can then implement adapters for multiple environment (e.g., a JavaScript web-browser environment, a Python environment, a Java environment).
The executing environment can then select the most suitable adapter among the available ones.

When comparing \textit{technology interfaces} to OOP primitives, we think of them as functional interfaces, as they only provide exactly one \texttt{execute}-function that has a set of parameters and a set of return values.
These functional interfaces may then be implemented by the adapters that can be injected into the executing service as a dependency, resembling the dependency inversion principle of OOP.

\subsubsection{Attribute Mappings}
\label{subsubsec:model:Mappings}
Since the \textit{attributes} provided as input to an \textit{operation} are not necessarily identical to those specified by an \textit{operation} or \textit{technology interface}, we need to map between these \textit{attributes}.
The \textit{Attribute Mapping} class provides such a mapping mechanism and therefore enables the reuse of \textit{operations} and \textit{technology interfaces}.
Figure~\ref{fig:operation-example} visualizes such a use case where the \textit{operation} (light blue) demands a ``contact'' \textit{attribute}, which is then transferred into the ``regexInput'' \textit{attribute} of the ``regex'' \textit{technology interface}.
Similarly, an element from the output of the ``regex'' \textit{technology interface} is extracted, transformed to the definition of an \textit{attribute} and used in another \textit{operation step}.
We recall that every \textit{attribute} conforms to a \textit{data type} specifying its syntax.
\textit{Attribute mappings} therefore must fulfill multiple roles also known in OOP:
\begin{itemize}
    \item \textbf{Defining constant values:} Not all \textit{attributes} of \textit{technology interfaces} need to be present in every FAIR-DO.
    Information such as the regex pattern extracting parts of an ORCiD-URL, the source code location of a Python script, or the setup necessary to run the script pertains to the \textit{operation} itself rather than to the FAIR-DO it operates on.
    Therefore, \textit{attribute mappings} support the specification of constant values within an \textit{operation step} by providing a ``value'' field.
    \item \textbf{Type casting:} Different \textit{attributes} often conform to different \textit{data types}.
    For example, the ``regexInput'' \textit{attribute} of the ``Regex'' \textit{Technology Interface} naturally conforms to an arbitrary String.
    We therefore down-cast the ``contact'' \textit{attribute}, that conforms to the syntax of an ORCiD-URL, to a String to use it with the ``Regex'' \textit{technology interface}.
    To make this mechanism work for both down- and up-casting, the executing system needs to validate the input values for the \textit{attribute mappings} against the \textit{data type} the output \textit{attribute} complies to.
    \item \textbf{Addressing items in an array:} In Figure~\ref{fig:operation-example}, there are two examples for the use of up-casting mechanisms, although they do not perform a 1-to-1 mapping, but rather an n-to-1 mapping.
    The ``regexOutput'' and ``returnValues'' \textit{attributes} have a cardinality greater than one, and can therefore be considered as an array of values.
    However, the ``extracted ORCiD'' and ``e-mail address'' \textit{attributes} have a cardinality of exactly one, which necessitates the \textit{attribute mappings} to select one element of their input arrays.
    The \textit{attribute mappings} refer to the input and output \textit{attribute} and specify the index of the element to be used.
    The validity has to be enforced by the executing system.
    \item \textbf{Providing templates for Strings:} String templating, the process of adding the contents of a variable to a predefined string, is well known from programming and also of relevance in our operation model.
    In our example in Figure~\ref{fig:operation-example}, we use this String template mechanism to insert the ``extracted ORCiD'' from the first \textit{operation step} into the ``runCommand'' \textit{attribute} that executes the Python script.
    The pattern of the insertion position (by default \textit{``\{\{input\}\}''}) can be changed to facilitate as many use cases as possible.
\end{itemize}

\section{Model Implementation}
\label{sec:Impl}
We introduce IDORIS as a prototypical implementation of our typing model that can be found on GitHub\footnote{\url{https://github.com/maximiliani/IDORIS}}.
The full name ``Integrated Data Type and Operations Registry with Inheritance System'' reflects the essential functionalities of our typing model described in Section~\ref{sec:Model}.
Technologically, IDORIS is a Spring Boot\footnote{\url{https://spring.io}} microservice, developed in Java 21\footnote{\url{https://java.com}}.
For storage, the graph database Neo4j\footnote{\label{foot:neo4j}\url{https://neo4j.com}} is used in combination with Spring Data Neo4j\footnote{\url{https://docs.spring.io/spring-data/neo4j/reference}} and Spring Data REST\footnote{\url{https://docs.spring.io/spring-data/rest/reference}}, making IDORIS capable to provide an automatically generated and fully HATEOAS-enabled RESTful-API for CRUD functionality, solely based on our model.
More advanced features that demand additional logic, such as resolving the inheritance hierarchy and retrieving available \textit{operations} for \textit{data types}, are exposed using traditional Spring Web MVC endpoints\footnote{\url{https://docs.spring.io/spring-framework/reference/web/webmvc.html}}.

\subsection{Graph database}
\label{subsec:Impl:Graph}
Due to the high inter-connectivity of our model, efficient querying of relationships between model components is essential.
This is a typical use-case for graph databases.
We chose Neo4j for its labeled-property graph model and integrability into the technology stack of IDORIS, which allows us to implement our typing model with only minor technical changes, thus enhancing the expressivity of the graph.
IDORIS uses these efficient in-database processing capabilities to find all \textit{operations} executable on an \textit{attribute} or transitively a \textit{data type}, to detect cycles in the graph, and to resolve inheritance hierarchies.
Furthermore, we can use graph algorithms for path finding, circle detection, and relationship querying provided by Neo4j directly inside our graph database.

\subsection{Rule-based validation and processing}
\label{subsec:Impl:Validation}
Since our model depends heavily on the correctness of user-provided information, IDORIS must validate this data both, syntactically and semantically.
Especially when realizing the inheritance mechanisms for \textit{atomic data types} and \textit{type profiles}, a validation mechanism that is able to act not only on individual entities but also on their contextual relationships is needed.
This feature is clearly beyond the capabilities of a JSON schema.

We realized a modular, rule-based approach for validation to enhance its maintainability.
This is accomplished by using the ``Visitor'' design pattern~\cite{gammaDesignPatternsElements1995} that separates logic from model classes in a highly modular fashion and is therefore often used, among others, for semantic validation and optimization inside compilers.
Each validation rule is implemented in a separate Visitor class, having a dedicated behavior for each model class it is called upon (e.g., \textit{Atomic Data Type}, \textit{Type Profile}, \textit{Operation}).
Visitors perform non-trivial validations of the inheritance hierarchy and of relations to other entities (such as \textit{attributes}).
This is primarily done through recursion, interaction with the accessor methods, and interaction with the graph database to ensure cross-entity consistency.
This approach ensures that the inheritance hierarchy is free of conflicts and circular dependencies.
In IDORIS, Visitors are currently only used for validation purposes, but are designed to solve future problems such as JSON schema generation or optimization algorithms.

\begin{figure*}[htbp]
\centerline{\includegraphics[width=7in]{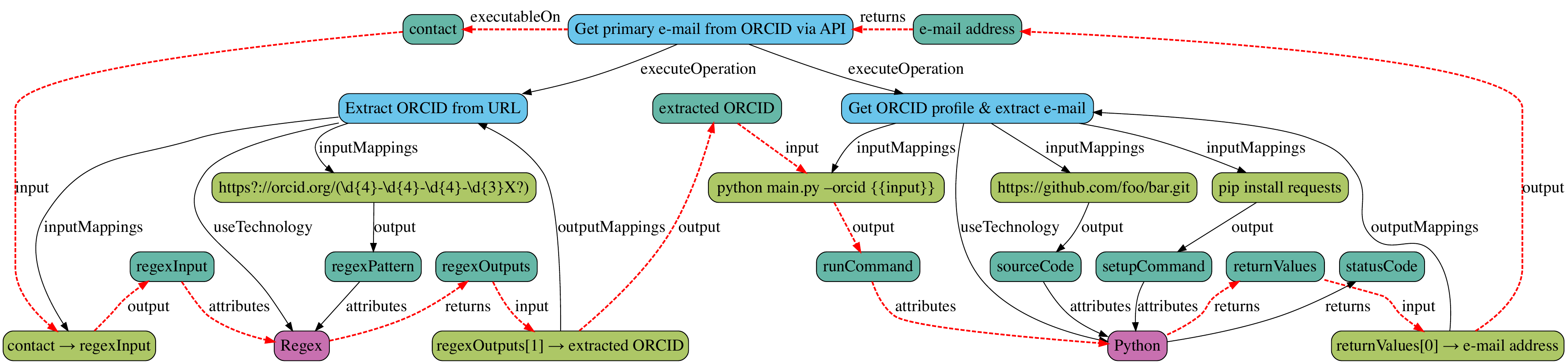}}
\caption{Excerpt from the Graph representing a FAIR-DO Operation stored in the database}
\label{fig:useCaseGraph}
\end{figure*}

\section{Evaluation and Discussion}
\label{sec:Evaluation}
\subsection{IDORIS-based Use-Case}
\label{subsec:Evaluation:UseCase}
We visualized our typing model in Figure~\ref{fig:typing-model}, defined our approach for type-associated FAIR-DO Operations in Subsection~\ref{subsec:model:ops}, and illustrated a running example in Figure~\ref{fig:operation-example}.
In this subsection, we will reuse this running example as a use-case and provide an excerpt of the actual graph data structure in Figure~\ref{fig:useCaseGraph} (as described in Subsection~\ref{subsec:Impl:Graph}) with all relevant nodes and relations that describe an \textit{operation}.
For better readability, we will only show a description of the contents of each node, omitting its properties.
The color-labels of the nodes in our labeled-property graph match those of Figures~\ref{fig:typing-model} and~\ref{fig:operation-example}: the \textit{operation} and its \textit{operation steps} are in light blue, \textit{attributes} in dark green, \textit{attribute mappings} in light green, and \textit{technology interfaces} in purple.
This graph can be retrieved by executing the simple Cypher-Query: \texttt{MATCH (n: Operation | AttributeMapping | Operation Step | TechnologyInterface | Attribute) RETURN n)}.
Therefore, we can argue that the graph directly represents the classes of our UML-based typing model in a semantically meaningful manner.

The semantically meaningful nodes and relations in our graph database can be exploited for more complex queries:
with respect to our example, the red relations form a circle in our graph, representing the flow of data within the ``Get primary e-mail from ORCiD via API''-\textit{operation}.
It is visible how the data flows from the ``contact'' \textit{attribute} through the \textit{attribute mapping} into the ``regexInput'' \textit{attribute}, that is processed by the ``Regex'' \textit{technology interface} whose output is then again transformed and inserted into a command starting a ``Python script'' that outputs the ``e-mail address''.
This circle is queried using the Cypher query: \texttt{MATCH (m1) WITH collect(m1) as nodes CALL apoc.nodes.cycles(nodes) YIELD path RETURN path)}.
The same query returns an additional circle for each \textit{operation step}, representing levels of abstraction inside an \textit{operation}.
This circle detection is useful for a future executing system to automatically parallelize processing and ensuring data is available when it is needed.
The concrete use case of this example can be simplified by using a platform-independent execution mechanism, such as Web Assembly (WASM)~\cite{webassemblyworkinggroupWebAssemblyCoreSpecification2019}, whose limitations are outside the scope of this work.
However, more complex use cases that need mechanisms for environment-specific execution, optimization, and access to resources that only native code can provide may use technologies such as Docker Containers and Python Scripts, demanding a more flexible modeling approach.
We do not intend to develop a ``universal programming language'', but instead to facilitate the variety of languages, frameworks, and tools that already exist with our technology-agnostic model for FAIR-DO Operations.

IDORIS must also ensure that no invalid cycles are introduced.
Examples for such unwanted circular dependencies include, but are not limited to: circles in the inheritance hierarchy, \textit{type profiles} using themselves as \textit{attributes}, and an \textit{operation} calling itself within an \textit{operation step}.
To avoid this, we used the path finding algorithms of Neo4j and created a rule for IDORIS' rule-based validator system (Subsection~\ref{subsec:Impl:Validation}).
In Listing~\ref{lst:ValidatorCode}, we show an excerpt of such a validator in Java-like pseudocode that also creates error messages via the API including severity level, message, and the entity of interest.
This feature of IDORIS assists users with detailed error messages when creating new elements and ensures data integrity.
By implementing separate validator classes for each rule, we enhanced the maintainability of our codebase through separation of concerns.
These validators can also be used to ensure semantic correctness, e.g., by ensuring no conflicts exist in the inheritance hierarchies of \textit{atomic data types} and \textit{type profiles}.
\begin{lstlisting}[language=Java, caption=Excerpt of the Acyclicity Validator role in pseudocode, label=lst:ValidatorCode]
public class AcyclicityValidator extends Visitor<ValidationResult> {
    private final Neo4jClient neo4jClient;

    // Visitor method to ensure an operation does not call itself
    public ValidationResult visitOperation (Operation operation) {
        return doesNotExecuteItself(operation);
    }

    // Visitor method to ensure it has no recursive dependency on itself
    public ValidationResult visitTypeProfile (TypeProfile typeProfile) {
        return ValidationResult.combine(
            doesNotInheritFromItself(typeProfile),
            doesNotUseItselfAsAttribute(typeProfile)
        ); 
    }
    [...]
    private ValidationResult doesNotInheritItself (DataType dataType){
        String query = "MATCH path = (n:DataType {pid: $nodePID})-[:inheritsFrom*1..]->(n) RETURN path LIMIT 1";

        // Query the path from the Neo4j database
        boolean hasCycle = neo4jClient.query(query)
            .bind(dataType.getPID()).to("nodePID")
            .fetch()
            .first()
            .isPresent();

            
        return hasCycle 
            ? new Error("Circular inheritance detected", dataType) 
            : new OK();
    }
 }
\end{lstlisting}

\begin{figure}[htbp]
\centerline{\includegraphics[width=3.5in]{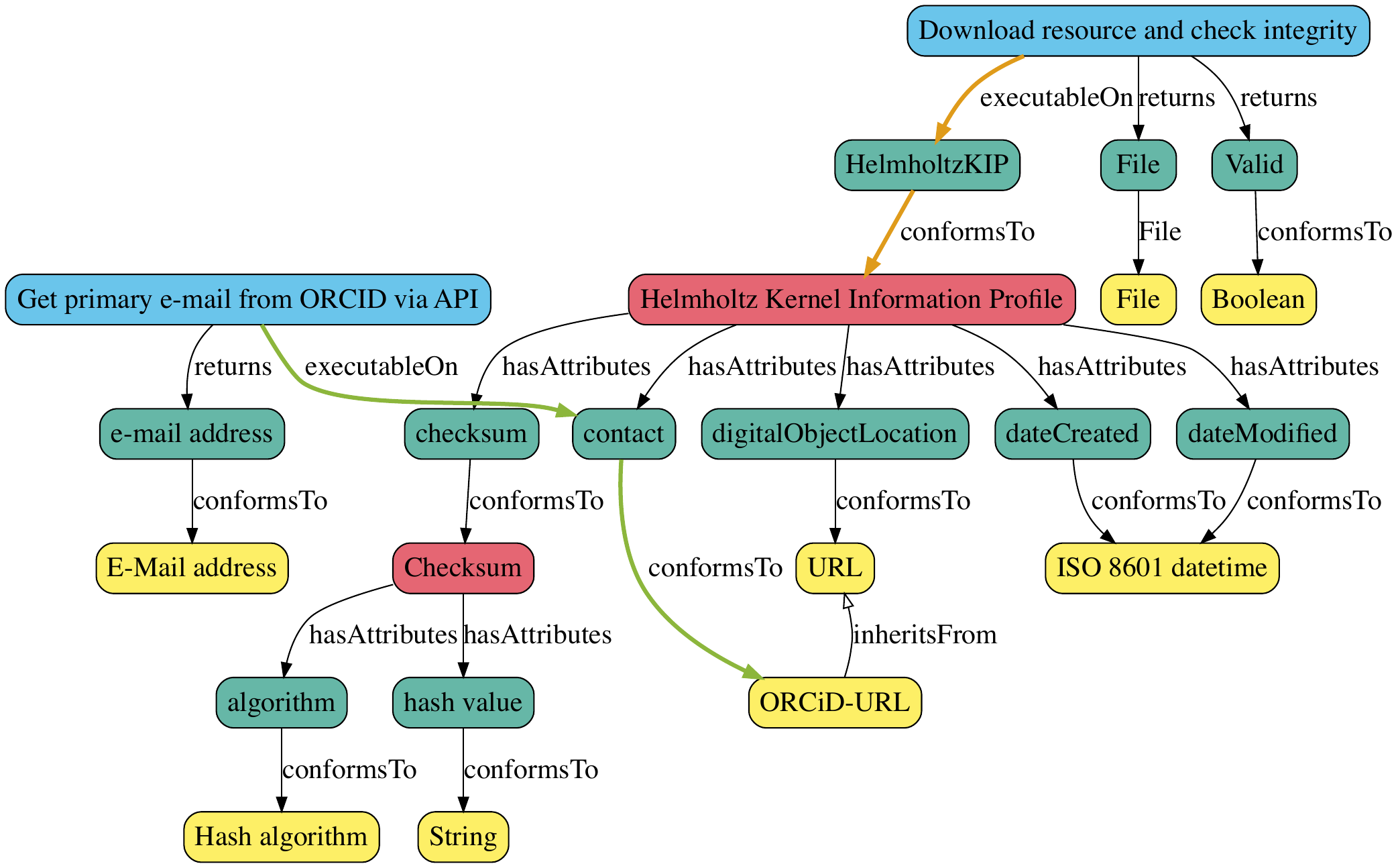}}
\caption{Excerpt from the Graph representing Data Types in the graph database}
\label{fig:useCaseDTGraph}
\end{figure}

Figure~\ref{fig:useCaseDTGraph} shows detailed examples of two application cases of \textit{Type Profiles} (described in Subsection~\ref{subsubsec:model:TypeProfile} and marked in red):
The ``Helmholtz Kernel Information Profile’’ \textit{type profile} is illustrated with excerpts containing selected \textit{attributes}.
This profile is used to describe the content of FAIR-DOs that adhere to it.
One of these \textit{attributes} (Subsection~\ref{subsec:model:Attributes}) conforms to the ``Checksum’’ \textit{type profile}. 
This profile is used to describe a complex JSON object embedded within a value in a FAIR-DO, utilizing the ``Helmholtz Kernel Information Profile’’. 
The resulting JSON object contains the hash and the algorithm, that generated the hash.
Furthermore, it shows an example for the inheritance of \textit{atomic data types} (Subsection~\ref{subsubsec:model:AtomicDataType}) by specifying that ``ORCiD-URL'' inherits from ``URL''.

Furthermore, Figure~\ref{fig:useCaseDTGraph} visualizes \textit{data types}, namely \textit{atomic data types} (yellow) and \textit{type profiles} (red), in addition to the \textit{attributes}, and \textit{operations} using the running example and an additional example operation ``Download Resource and Check Integrity''.
We use these to elaborate on the fulfillment of conceptual association mechanisms for operations and FAIR-DOs according to ``Record Typing'', ``Profile typing'' and ``Attribute typing'' (as introduced in Section~\ref{sec:SOTA}).
Since our \textit{operations} are assigned with PIDs, we support referencing them from FAIR-DOs, enabling ``Record typing''.
We realize both ``Profile typing'' and ``Attribute typing'' by specifying a single \textit{attribute} an \textit{operation} is executable on.
For ``Profile typing'', this \textit{attribute} conforms to a \textit{type profile} (orange path in Figure~\ref{fig:useCaseDTGraph}).
For ``Attribute typing'', this \textit{attribute} can conform to any \textit{data type}, namely \textit{atomic data types} or a \textit{type profile} (light green path in Figure~\ref{fig:useCaseDTGraph}).
However, we decided to not adopt the duck typing variant of ``Attribute Typing'' by allowing only exactly one \textit{attribute} an \textit{operation} is executable on.
Hence, all FAIR-DO operation association mechanisms are (at least partially) supported by our model and implemented in IDORIS.

\subsection{Comparison to Existing Data Type Registry Models} 
\label{subsec:Evaluation:ePIC-DTR}
We designed our integrated typing model based on the concepts and development of the ePIC and EOSC DTRs~\cite{schwardmannAutomatedSchemaExtraction2016a}.
The core concepts — defining simple value syntax and structuring complex values — remain unchanged.


In the ePIC and EOSC DTRs, \textit{PID-BasicInfoTypes} and \textit{PID-InfoTypes} are sometimes called Data Types for simplicity.
\textit{PID-BasicInfoTypes} for the syntax of simple values are modeled by the \textit{Atomic Data Type} class (Subsection~\ref{subsubsec:model:AtomicDataType}).
To define complex structures, our model combines the \textit{PID-InfoTypes} (for complex JSON values inside a FAIR-DO) and the \textit{KernelInformationProfiles} (for the structure of FAIR-DOs) with the \textit{Type Profile} class (Subsection~\ref{subsubsec:model:TypeProfile}).
As a new approach, \textit{Atomic Data Types} and \textit{Type Profiles} themselves are abstracted by the \textit{Data Type} class, which reduces redundancies and provides a strong definition of the term ``data type'' within our model, enhancing its semantic clarity.
The newly introduced inheritance mechanisms for both, \textit{Atomic Data Types} and \textit{Type Profiles} promote reusability of their instances and already allows for basic polymorphic behavior through subtyping, facilitating reuse of the association between \textit{data types} and \textit{operations}.
This relates to the model's ability to specify machine-actionable \textit{operations} that are associated with \textit{attributes} (and transitively \textit{data types}), enabling type-associated FAIR-DO Operations.
Unlike \textit{PID-BasicInfoTypes}, our \textit{Atomic Data Types} do not contain fields for specifying measurement units or categories, and exclude other rarely used or undocumented fields, to enhance the semantic clarity of the typing model.
For the same reason, the former \textit{SubSchemaRelation} for PITs and KIPs~\cite{schwardmannAutomatedSchemaExtraction2016a} was split into a validation policy and a flag indicating whether additional attributes are allowed.

Unlike the ePIC and EOSC DTRs, IDORIS is not based on JSON schema.
Instead, we use a graph database, ideal for storing highly connected entities and executing graph algorithms to query the inheritance hierarchy of data types and for finding \textit{operations} executable on an \textit{attribute} or \textit{data type}.
We also provide a more capable rule-based validation logic that is able to validate more than just the syntax of single entities.
This way, we can describe type-associated operations and ensure the quality of information stored in IDORIS.
We decided against an RDF-based system due to its limited integration into the Spring framework, higher modeling complexity with triples, and steeper learning curve.
Although RDF-based graph databases, such as Apache Jena\footnote{\url{https://jena.apache.org}}, offer potential advantages, including an easier integration into knowledge graphs, the ability to reuse terms from ontologies, and possible support for federated queries beyond single systems, concrete use-cases benefiting from these features have not yet been identified.

\section{Conclusion and Future Work}
\label{sec:Conclusions}
Our integrated model significantly contributes to the long-term vision of machine-actionable FAIR-DOs by introducing type-associated FAIR-DO Operations that are well described in a technology-agnostic and highly reusable manner.
We improved upon the models used in existing DTRs by integrating inheritance mechanisms, streamlining the components of the type system, and providing mechanisms that associate data types to FAIR-DO Operations.
To demonstrate the feasibility and capabilities of our model, we developed IDORIS, a prototype for a next-generation of DTRs that can accommodate data types, technology-agnostic FAIR-DO Operations, and perform computations to associate them.
In addition, IDORIS incorporates a robust validation system that provides detailed feedback to ensure data integrity.
Our model and IDORIS therefore provide a foundational typing infrastructure improving interoperability, reusability, and automation in research data management, and enabling future work towards executing FAIR-DO Operations.

Upcoming research includes developing execution components for FAIR-DO Operations including adapters for widely adopted technologies, enhancing system scalability, and integrating federated DTR instances.

\bibliographystyle{IEEEtran}
\bibliography{IDORIS-paper}

\end{document}